\documentclass[pra]{revtex4}
\usepackage{amssymb}
\usepackage{amsmath}
\usepackage{graphicx}

\begin{document}
\title{Phase Model with Feedback Control for Power Grids in Kyushu Region}
\author{Tatsuma Matsuo and Hidetsugu Sakaguchi}
\affiliation{Department of Applied Science for Electronics and Materials,
Interdisciplinary Graduate School of Engineering Sciences, Kyushu
University, Kasuga, Fukuoka 816-8580, Japan}
\begin{abstract}
A phase model with feedback control is studied as a dynamical model of power grids. As an example, we study a model network corresponding to the power grid in the Kyushu region. The standard frequency is maintained by the mutual synchronization and the feedback control. Electric failures are induced by an overload. 
We propose a local feedback method in which the strength of feedback control is proportional to the magnitude of generators. We find that the electric failures do not occur until the utilization ratio is close to 1 under this feedback control.  We also find that the temporal response for the time-varying input power is suppressed under this feedback control. We explain the mechanisms using the corresponding global feedback method.
\end{abstract}
\maketitle
\section{Introduction}
The synchronization in coupled oscillators has intensively been studied~\cite{rf:1,rf:2}.  One of the recent topics is the synchronization in various types of networks~\cite{rf:3,rf:4,rf:5}. The network structure itself is intensively investigated in physics. There are some reports on the network structure of power grids.  Barab\'asi and Albert reported that a power grid in the western United States is a scale-free network~\cite{rf:6}; however, some other authors reported that the link number distribution of transmission lines is an exponential distribution~\cite{rf:7,rf:8}. 
A serious issue in power grids is a large-scale electric failure called  blackout~\cite{rf:9}. Blackout or cascade failure in power grids is disastrous in the modern society. Since the alternating current (AC) system is used in power grids, not only the network structure but also electric potentials and phase variables are important for describing the behavior of power grids quantitatively. For example, the power supply to consumers becomes impossible, when the power demand is beyond the threshold. This is called the voltage collapse. 
Another important phenomenon in power grids is the step out, in which the frequency of a generator  deviates from the standard frequency such as 50 or 60 Hz owing to the very large power demand. These electric failures can lead to a large-scale blackout.

Several authors proposed some coupled oscillator models for describing the dynamical behaviors of power grids~\cite{rf:10,rf:11,rf:12}.  We  proposed a phase model with feedback control to maintain the standard frequency, and showed a cascade failure in a square lattice and a scale-free network~\cite{rf:13}. In this study, we numerically investigate the phase model in a network corresponding to the actual power grid in the Kyushu region.    

The network is composed of $N=97$ nodes of power stations and transformer substations and the transmission lines connecting the nodes. The total number of power stations in our model is $N_g=15$, which are denoted by squares in Fig.~1(a). There are various magnitudes of power stations or generators in the Kyushu region. For example, the maximum output power of the "Genkai" nuclear power plant is $3.5\times 10^6$ kW and the maximum output powers of the "Ohmuta" solar power plant is $3\times 10^3$ kW. The sum of the maximum output powers of the 15 power stations is $17.3\times 10^6$ kW.  In our previous work, we studied model systems in a square lattice and a scale-free network, where the maximum output powers of generators were assumed to be the same. In real power grids, the maximum output powers of generators are very widely distributed. In this paper, we will propose a simple and rather effective feedback control method applied in such heterogeneous systems.    
The $N_t=82$ nodes denoted by the plus sign in Fig.~1 are transformer substations.  The links between the generators and the transformer substation and between two transformer substations are transmission lines. There are two types of transmission lines, i.e., 500kV transmission lines and 275kV transmission lines. 
Transmission lines of lower voltage are extended from the transformer substations to local consumers, but they are not shown in Fig.~1.  In our model, the lower-level power grids are effectively treated as consumed electric power at the transformer substations.   
\begin{figure}[tbp]
\begin{center}
\includegraphics[height=4.cm]{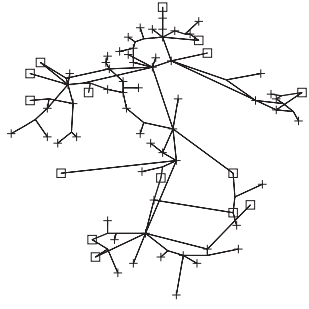}
\end{center}
\caption{Power grid model in the Kyushu region. The power stations are denoted by squares and the transformer substations are denoted by the plus sign. }
\label{fig1}
\end{figure}
\section{Phase Model of Power Grids and Feedback Control}
We explain the phase model of power grids proposed in the previous paper~\cite{rf:13}. In synchronous generators, the rotational motion of a turbine is driven by mechanical power such as the steam power generated by burning fuel.  The rotational motion is directly transformed into the sinusoidal alternating current.  The phase variable of the alternating current for the $i$th generator is denoted as $\theta_i$. We use another phase variable, $\phi_i=\theta_i-\Omega_0 t$, where $\Omega_0$ is the standard frequency 50 or 60 Hz, because the phase $\phi_i$ is more convenient for denoting the deviation from the standard frequency. The equation for $\phi_i$ is assumed to obey  
\begin{equation}
\frac{d^2\phi_i}{dt^2}=-D\frac{d\phi_i}{dt}+W_i-P_{gi},
\end{equation}
where $D$ is the damping constant, $W_i$  denotes the input power to the $i$th generator, and $P_{gi}$ is the output power from the $i$th generator. 
The output power $P_{gi}$ is expressed as
\begin{equation}
P_{gi}=\sum_jE_iE_jY_{ij}\sin(\phi_i-\phi_j),
\end{equation}
where $E_i$ is the electric potential at the $i$th generator, and $Y_{ij}$ denotes the admittance between the $i$th and $j$th nodes. Here, we have assumed that the resistance $R_{ij}$ between the $i$th and $j$th nodes is neglected for simplicity.  The summation in Eq.~(2) is taken for all $j$ nodes linked to the $i$th generator.  The electric potential $E_i$ and the admittance $Y_{ij}$ are normalized using base values of the electric potential such as 500 or 275 kV and an electric power of 1000 MW. The admittance of the transmission line depends on the distance between the two nodes, which is evaluated from the map shown in Fig.~1. 
In this paper, the output voltage $E_i$ for generators is assumed to take a constant value of 1 in the normalized unit. 

The generated electric power is transmitted through the transfer substations to  consumers. We assume that a constant quantity of electric power is consumed under the $i$th transfer substation.  The consumed electric power at the $i$th node is expressed by the complex number $P_{ei}+iQ_{ri}$, where $P_{ei}$ and $Q_{ri}$ are the effective power and  reactive power, respectively. The complex power $P_{ei}+iQ_{ri}$ is expressed as $E_{i}e^{i\phi_i}I_i^{*}$, where $E_i$ is the electric potential,  $I_i$ is the current supplied to the $i$th node, and $^*$ implies the complex conjugate.  From the conservation law of current, $I_i$ is expressed as
\begin{equation}
I_i=\sum_j-iY_{ij}\left (E_je^{i\phi_j}-E_ie^{i\phi_i}\right ).
\end{equation}
The powers $P_{ei}$ and $Q_{ri}$ are therefore given by
\begin{eqnarray}
P_{ei}&=&\sum_jY_{ij}E_iE_j\sin(\phi_j-\phi_i),\\
Q_{ri}&=&\sum_jY_{ij}\{E_iE_j\cos(\phi_j-\phi_i)-E_i^2\}.
\end{eqnarray}
 The summation is taken for all nodes $j$ linked to the $i$th node.

The feedback control is applied to maintain the standard frequency.
One of the simplest model of feedback control is   
\begin{equation}
\frac{dW_i}{dt}=-\gamma_i \frac{d\phi_i}{dt},
\end{equation}
where $\gamma_i>0$ denotes the strength of the feedback control. 
If $d\phi_i/dt$ is positive or the frequency is higher than the standard value, the input torque power $W_i$ decreases. Conversely, if $d\phi_i/dt$ is negative or the frequency is lower than the standard value, the input torque power increases. If the feedback control works well, the frequency $d\phi_i/dt$  is maintained at zero, and the entire power system including the generators and loads works at the standard frequency $\Omega_0$. 
The above feedback control method is a local feedback, because the input power to the $i$th generator is controlled by the local value $d\phi_i/dt$ of the same generator. The actual power system is regulated not only by such a local feedback but also by the global control from the control center using all the information of the network. A relation expressing such a nonlocal feedback control is written as   \begin{equation}
\frac{dW_i}{dt}=-\sum_{j=1}^{N_g}\gamma_{i,j}\frac{d\phi_j}{dt},
\end{equation}
where $\gamma_{i,j}>0$ denotes the effect from the $j$th generator to the $i$th generator. 

If the power demand expressed by $P_{gi}$ is too large, the input power $W_i$ cannot respond to the excessive demand. If generators are overloaded, they might break down.  To avoid such breakdown, generators are stopped, or the overall synchronization would be lost. That is, a loss of synchronization would occur, which is called the step out. If some generators stop, the load to other generators increases and some other generators are overloaded successively. This leads to a cascade failure. 
 To express such a situation, we assume a simple rule that $E_i$ is set to zero if $W_i$ is beyond the critical value $W_{ci}$, where $W_{ci}$ can be interpreted as the maximum output power of the $i$th generator.  The magnitudes of $W_{ci}$ are very widely distributed in our network, because various types of generators work in the Kyushu region. If $E_i(t)$ is set to zero, the function of the generator is lost. This is equivalent to the removal of the generator node $i$ from the network. 

In our model, the sum of the maximum output powers of the 15 power stations is fixed at $SW_{c}=\sum_{i=1}^{N_g} W_{ci}$. Similarly, the sum of the consumed effective powers is expressed as $SP=\sum_i^{N_t}P_{ei}$, and the sum of the reactive powers is $SQ=\sum_i^{N_t}Q_{ri}$. In our model power grid, we assume that the consumed power $P_{ei}$ at the $i$th node is  proportional to the population $p_i$ in the area around the $i$th node, which is also estimated from the map, and the ratio of $Q_{ri}$ to $P_{ei}$ is constant. The ratio of the total consumed power to the total maximum output power of generators is called the utilization ratio $r$. 
Using these assumptions, $P_{ei}$ and $Q_{ri}$ are expressed as 
\begin{equation}
P_{ei}=\left (\frac{p_i}{\sum p_i}\right )(r\cdot SW_c), \;Q_{ri}=qP_{ei},
\end{equation}
where $0<r<1$ is the utilization ratio of the electric power, and $1/\sqrt{1+q^2}$ denotes the power factor.  In our setting, the effective  power of $r\cdot SW_c$ is consumed in the entire region and the effective power for each area around the $i$th node is proportional to the population of the area. The utilization ratio $r=1$ implies that the maximum output power from the generators is completely consumed in the entire region.

The phase $\phi_i$ and electric potential $E_i$ are determined by solving the coupled equations Eqs.~(4) and (5) for the assigned values of $P_{ei}$ and $Q_{ri}$.
These coupled equations can be numerically solved by an iterative method. 
If $P_{ei}$ and $Q_{ri}$ are sufficiently small,  there are two stationary solutions: one is stable and the other is unstable. However, if the effective power $P_{ei}$ is gradually increased, the stable solution $E_i$ decreases and the unstable one increases.  At a critical value of $P_{ei}$, the stable and unstable solutions merge and disappear. There are no stationary solutions beyond the critical value.  This phenomenon is called the voltage collapse. If the voltage collapse occurs, the voltage $E_i$ decreases rapidly. Generally, the voltage collapse occurs more easily if the reactive power $Q_{ri}$ is large.  If the voltage collapse occurs, there is no stationary solution, or the power supply to the $i$th node becomes impossible. In our simple model,  $E_i$ is set to zero when $E_i(t)$ decreases to zero.  This is equivalent to removing the node $i$ from the network. The two removal rules correspond to the two types of electric failures, i.e., the step out and voltage collapse. 

We numerically study electric failures in the model network using the phase model with feedback control, as the utilization ratio $r$ is increased. The parameter $D$ is fixed to be 1. The initial values of $W_{i}$, $\phi_i$, and $d\phi_i/dt$ are set to 0 in our numerical simulation.

\begin{figure}[tbp]
\begin{center}
\includegraphics[height=3.5cm]{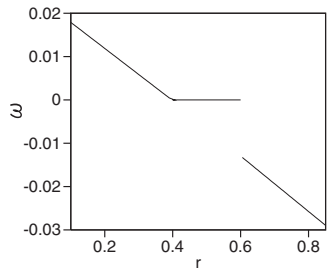}
\end{center}
\caption{Average frequency $\omega$ as a function of the utilization ratio $r$.}\label{fig2}
\end{figure}
Firstly, we show synchronization-desynchronization transitions when only one generator is controlled by the local feedback, as expressed by Eq.~(6). 
Figure 2 shows the average frequency $\omega=\sum_{i=1}^{N_g}\langle d\phi_i/dt\rangle/N_g$, where $\langle \cdots \rangle$ denotes the long-time average when $\gamma_1=0.5$ and $\gamma_i=0$ for the other generators. Here, the first generator is the generator at Genkai with the largest maximum output power. That is, the feedback control is set only for the first generator. The input power $W_i$ for $i\ne 1$ is fixed to be $0.5W_{ci}$.  The parameter $q$ is set to be 0; therefore, the power factor is 1.  
When the utilization ratio $r$ is between 0.4 and 0.6, the average frequency is maintained to be 0. The average frequency of all the generators is 0 by the mutual synchronization. The standard frequency is realized in this power grid owing to the local feedback control and  mutual synchronization.
When $r$ is smaller than 0.4, the torque $W_1$ of the first generator becomes 0. The total input power $\sum_{i\ne 1}W_i$ is larger than the power demand $r\cdot SP$, and the frequency of generators becomes larger than 0 continuously.  On the other hand, the input power $W_1$ of the first generator reaches its maximum  $W_{c1}$, and the first generator stops when $r$ is larger than 0.6.  The total input power $\sum_{i\ne 1}W_i$ is insufficient for the power demand. Then the frequency of generators decreases from the standard value discontinuously, because the input power of the first generator decreases to 0 discontinuously.  
\section{Feedback Control in Proportion to the Magnitude of Generators}
It is desirable to maintain the standard frequency up to $r=1$, where the maximum output of all generators is completely consumed by consumers; however, electric failure tends to occur easily when $r$ is close to 1.   

We propose a local feedback control method of the following form as one efficient feedback control method:  
\begin{equation}
\frac{dW_i}{dt}=-\gamma_i\frac{d\phi_i}{dt}=-\gamma \frac{W_{ci}}{SW_c}\frac{d\phi_i}{dt}.
\end{equation}
It is a feature of this method that the strength of the feedback is proportional to the maximum output power of the generator.  The method works fairly well as shown in the following, even if the maximum output powers of generators are very widely distributed. 

Figure 3(a) shows the number ratio $R_s$ for $\gamma=0.3$. The number ratio of the step out, $R_s$, is zero up to $r=0.955$, and increases stepwise to 0.133 at $r=0.96$ and to 0.8 at $r=0.99$. This implies that the power grid is rather stable.  A cascade failure is observed at $r=0.99$, because the step out of some generators makes the surviving generators overloaded, and it leads to the step out of the surviving ones. Figures 3(b) and 3(c) show the power grids at $t=30$ and 180, respectively.  At $t=30$, a few generators with small maximum output power  exhibited the step out and links from the nodes disappeared. 
The links from the step-out generators disappeared successively, and only three generators survived at $t=180$ after the cascade failure. The surviving generators are the ones whose maximum output power was large. 
\begin{figure}[tbp]
\begin{center}
\includegraphics[height=4.cm]{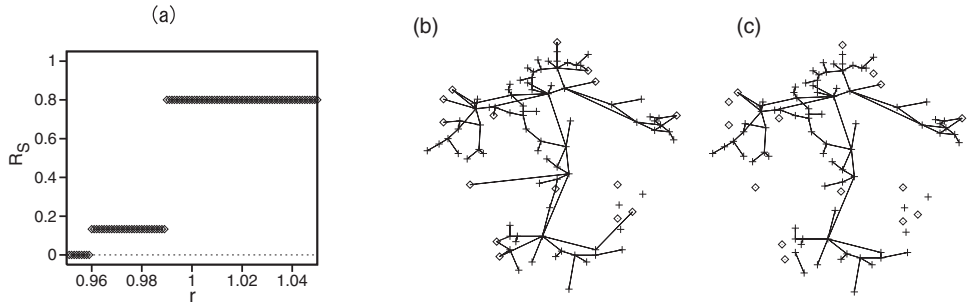}
\end{center}
\caption{(a) Number ratio $R_s$ as a function of $r$ for the local feedback model. Cascade failure in a local feedback model of Eq.~(9) at (b) $t=30$, and (c) $t=180$ for $r=0.99$.}
\label{fig3}
\end{figure}
\begin{figure}[tbp]
\begin{center}
\includegraphics[height=3.5cm]{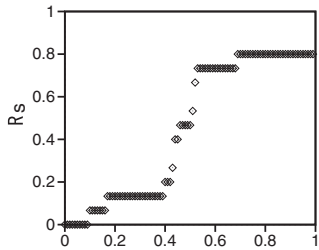}
\end{center}
\caption{Number ratio $R_s$ as a function of $r$ for the local feedback model of $\gamma_i=0.1$.}
\label{fig4}
\end{figure}
\begin{figure}[tbp]
\begin{center}
\includegraphics[height=4.cm]{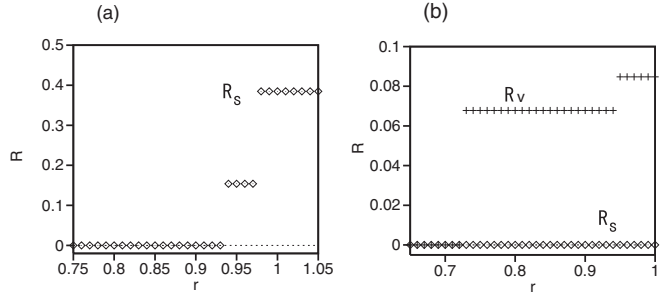}
\end{center}
\caption{Number ratio $R_s$ (rhombi) of the step out and the number ratio $R_v$ of the voltage collapse as functions of $r$ for (a) $q=1$ and (b) $q=2$.}
\label{fig5}
\end{figure}

To show the effectiveness of the local feedback method expressed by Eq.~(9), we consider a system where $\gamma_i=0.1$ for all the generators. That is, the strength of the feedback is independent of the maximum output power of the generator. As shown in Fig.~4, $R_s$ becomes nonzero for $r>0.1$. That is, the step out occurs  for a very small $r$. This is because generators with a small maximum output power tend to exhibit the step out easily even for a rather small $r$.  An efficient feedback control method needs to be devised in such a way that the electric failure does not occur up to $r\sim 1$. The local feedback method expressed by Eq.~(9) is one of the methods.

We have furthermore investigated the effect of a nonzero $q$ for the local feedback method expressed by Eq.~(9) with $\gamma=0.5$. The voltage collapse is generally considered to occur easily for a large $q$. 
Figure 5(a) shows the number ratio $R_s$ at $q=1$. The step out does not occur up to $r=0.94$. The voltage collapse does not occur below the critical value of $r=0.94$.  This implies that the step out is avoided even for a nonzero $q$. 
Figure 5(b) shows the number ratio $R_{s}$ for generators and the number ratio $R_v$  at which the voltage collapse occurred at $q=2$. 
The voltage collapse occurs first at $r=0.73$ at this large $q$. If the voltage collapse occurs at a certain node, the power demand decreases at this node, and the load to generators decreases. As a result, the step out does not occur and $R_s$ remains zero up to $r=1$.   The voltage collapse occurs first in the case of $q\ge 1.5$, if the other parameters are the same as those shown in Fig.~3(a). 
\begin{figure}[tbp]
\begin{center}
\includegraphics[height=3.5cm]{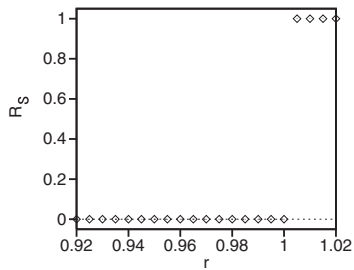}
\end{center}
\caption{Number ratio $R_S$ of the step-out generators as a function of $r$ in the global feedback model. }
\label{fig6}
\end{figure}

To understand the reason why the step out is avoided in the local feedback model expressed by Eq.~(9), even for a relatively large $r$, we consider a corresponding global feedback model of the form 
\begin{equation}
\frac{dW_i}{dt}=-\gamma \frac{W_{ci}}{SW_c}\left (\frac{1}{N_g}\sum_{j=1}^{N_g}\frac{d\phi_j}{dt}\right ),
\end{equation}
where $\gamma$ denotes the strength of the global feedback control. 
In this method, the feedback control is proportional to  the maximum output power $W_{ci}$ of each generator, which is the same as the local feedback model expressed by Eq.~(9). This feedback control is a global feedback in that the average $d\phi_j/dt$ of all the generators is used for controlling the $i$th generator.  
Because the initial values of $W_i$ are set to zero and $(1/N_g)\sum_{j=1}^{N_g}(d\phi_j/dt)$ takes the same value for all $i$'s, $W_{i}(t)$ is proportional to $W_{ci}$ for any time in this feedback control method.  
All the generators therefore reach the maximum output power $W_{ci}$ simultaneously when $r$ is increased. Note that the summation of $P_{gi}$ in Eq.~(2) with respect to all the nodes of generators is equal to the summation of $P_{ei}$ in Eq.~(4) with respect to all the nodes of transfer substations. Then, the summation of Eq.~(1) with respect to $i$ is reduced to 
\begin{equation}
\sum_{i=1}^{N_g}\frac{d^2\phi_i}{dt^2}=\sum_{i=1}^{N_g}\left (-D\frac{d\phi_i}{dt}+W_i\right )-\sum_{i=1}^{N_t}P_{ei}.
\end{equation}  
Because $d^2\phi_i/dt^2=0$ and $d\phi_i/dt=0$ in the stationary state while maintaining the standard frequency, the following equation is satisfied:  
\begin{equation}
\sum_{i=1}^{N_g}W_i=\sum_{i=1}^{N_t}P_{ei}.
\end{equation}
That is, the total input power is equal to the total consumed effective power. This is because the loss at the transmission lines is assumed to be zero. 
When $W_i$ becomes $W_{ci}$ for all the generators, 
\[\sum_{i=1}^{N_g}W_i=\sum_{i=1}^{N_t}P_{ei}=\sum_{i=1}^{N_g}W_{ci},\]
which implies that $r$ is equal to 1. 
This implies that the standard frequency is maintained up to $r=1$ in this global feedback model. 

We have checked it by the direct numerical simulation. Figure 6 shows the number ratio $R_s$ of generators that exhibited the step out for $\gamma=0.3$. 
$R_s$ is 0 up to $r=1$ and jumps to 1 for $r>1$. 
It is crucial that the strength of the feedback control for each generator is proportional to the maximum output power $W_{ci}$ in the feedback model to avoid the step out up to $r=1$. The local feedback model exoressed by Eq.~(9) and the global feedback model expressed by Eq.~(10) are different; however, the robustness against the step out for a large $r$ seems to originate from the same mechanism. 
The global feedback model can be interpreted as a kind of mean-field approximation for the local feedback model. 

\begin{figure}[tbp]
\begin{center}
\includegraphics[height=4.cm]{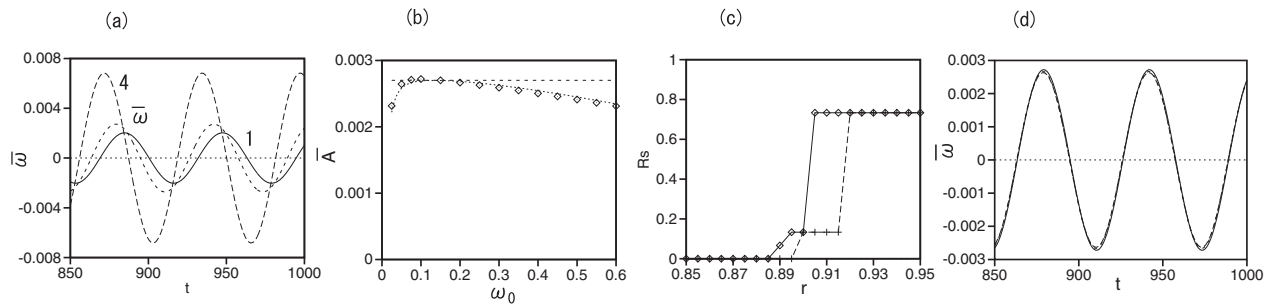}
\end{center}
\caption{(a) Time evolutions of $d\phi_1/dt$ (solid curve), $d\phi_4/dt$ (dashed curve), and $\bar{\omega}(t)$ for $\omega_0=0.1$ (dotted curve), when $W_4(t)=0.8W_{c4}\{1+\cos(\omega_0 t)\}/2$ with $\omega_0=0.1$. (b) Relationship between the amplitude $\bar{A}$ and $\omega_0$. The dashed line is $\bar{A}=0.0027$, and the dotted line is given by Eq.~(14). (c) Number ratio $R_s$ as a function of $r$ for the periodic input power (rhombi) and  constant power (plus). (d) The average frequency $\bar{\omega}(t)$ in the local feedback model (solid curve) and that in the global feedback model (dashed curve).}
\label{fig7}
\end{figure}
\section{Robustness against Temporal Fluctuations}
In this section, we study the effect of the temporal fluctuation in the input power $W_i(t)$, which cannot be controlled. The output of a thermal power station or a nuclear power station can be controlled by adjusting the degree of fuelling. However, the output power of a solar power station or a wind power station is difficult to control and fluctuates with time. 
As a model study, a numerical investigation of a case in which the input power of one generator changes with time and the other generators are controlled expressed by the local feedback of Eq.~(9) is carried out. As an example, the input power of the fourth generator is assumed to change as $W_4(t)=AW_{c4}\{1+\cos(\omega_0 t)\}/2$ with $A=0.8$ and $\omega_0=0.1$.
Figure 7(a) shows the time evolutions of $d\phi_1/dt$ (solid curve), $d\phi_4/dt$ (dashed curve), and $\bar{\omega}(t)=(1/N_g)\sum_{j=1}^{N_g}d\phi_j/dt$.   The other parameters are set to be  $q=0$,  $\gamma=0.3$, and $r=0.85$. The instant  frequency $d\phi_i/dt$ and its average $\bar{\omega}(t)$ oscillate owing to the periodic change in the input power $W_{4}(t)$. The amplitude of the oscillation for $i=4$ is larger than that for $i=1$, but it is much smaller than $0.4W_{c4}\sim 0.0408$. We have investigated the dependence of 
the amplitude $\bar{A}$ of the oscillation of $\bar{\omega}(t)$ on $\omega_0$. Figure 7(b) shows the relationship between $\bar{A}$ and $\omega_0$. 
The dashed line indicates $0.4W_{c4}/15\sim 0.0027$. The amplitude $\bar{A}$ changes with $\omega_0$; however, the dependence is not very large and it can be approximated at $0.4W_{c4}/15$. 
This result suggests that the temporal variation in the input power is averaged out by the mutual synchronization among generators. If the temporal variation in $\bar{\omega}(t)$ is too large, various harmful effects appear in the electric systems.  It is desirable that the oscillation amplitude in the entire power grid can be reduced by the mutual synchronization.  

Figure 7(c) shows the number ratio $R_s$ as a function of $r$ for a periodic input power $W_4(t)=0.8W_{c4}\{1+\cos(0.1 t)\}/2$ (rhombi) and a constant input power $W_4=0.4W_{c4}$ (plus). The step out appears at a slightly smaller $r$ for the periodic input than for the constant input which is the same as the temporal average of the periodic input, probably because the power supply is insufficient at a time of $W_4(t)\sim 0$.

To understand the response against the temporal fluctuation in the local feedback model, we again consider the global feedback model. That is, the periodic input $W_i(t)=AW_{ci}\{1+\cos(\omega_0 t)\}/2$ for $i=4$, $A=0.8$, and $\omega_0=0.1$ is applied to the global feedback model with $\gamma=0.3$. Figure 7(d) shows a comparison of $\bar{\omega}(t)$ for the local feedback control (solid curve) with that for the global feedback control (dashed curve). Good agreement is observed. 

For the global feedback control method, the average value $\bar{\phi}=\sum_{j=1}^{N_g}\phi_j/N_g$ obeys 
\begin{equation}
\frac{d^2\bar{\phi}}{dt^2}=-D\frac{d\bar{\phi}}{dt}-\frac{\gamma}{N_g}\left (1-\frac{W_{c4}}{SW_c}\right )\bar{\phi}-\frac{1}{N_g}\sum_i^{N_t}P_{ei}+\frac{1}{2N_g}AW_{c4}\{1+\cos(\omega_0 t)\},
\end{equation}  
because Eq.~(10) and $W_i=-\gamma (W_{ci}/SW_c)\bar{\phi}$ are satisfied in the global feedback model. 	 
This equation is equivalent to the equation of motion of the forced harmonic oscillator. The amplitude $\bar{A}$ of the oscillation of $\bar{\omega}(t)=d\bar{\phi}/dt$ is given by 
\begin{equation}
\bar{A}=\frac{AW_{c4}\omega_0}{2N_g\sqrt{\{-\omega_0^2+\gamma(1-W_{c4}/SW_c)/N_g\}^2+D^2\omega_0^2}}.
\end{equation}
If $\gamma/N_g$ and $\omega_0$ are sufficiently small, $\bar{A}$ is approximately given by $AW_{c4}/(2N_gD)$. This estimate of the amplitude of the oscillation is denoted by the  dashed line of $0.4W_{c4}/15=0.0027$ in Fig.~7(b). The dotted line in Fig.~7(b) denotes the curve expressed by Eq.~(14). The dependence of $\bar{A}$ on $\omega_0$ for the local feedback model is fairly well approximated by Eq.~(14) for the global feedback model. That is, the response for the temporal fluctuations in the local feedback system is fairly well approximated by that in the corresponding global feedback system. We do not show the result, but a similar type of periodic response was observed in the case that the effective power $P_{ei}$ changes with time periodically. 
 
\section{Summary and Discussion}
We have numerically studied a model power grid corresponding to the power grid in Kyushu region with a phase model. The maximum output powers of generators are very widely distributed in real power grids. We have found that a local feedback control expressed by Eq.~(9) is simple and rather efficient even in such heterogeneous systems. In the local feedback method, the strength $\gamma_i$ of the feedback control is set to be proportional to the maximum output power of the generator. It might be natural that the change rate of the torque $W_i$ is set to be proportional to the magnitude of the generator, but this is not trivial and a new finding that the step out does not occur up to $r\sim 1$ in this local feedback method. We have further studied the effect of temporal fluctuations in uncontrollable generators. The temporal fluctuations are suppressed by the mutual synchronization in the network system. The local feedback systems seem to be fairly well approximated in the corresponding global feedback systems, which can be interpreted as a mean-field approximation to the local feedback model.  
We have explained  the reason why the step out does not occur up to $r=1$ as well as the mechanism of the suppression of the temporal fluctuations in the global feedback systems. 

The phase model with the feedback control is a simple and useful model for analyzing the dynamical behavior of a power grid qualitatively; however,  it should be noted that more complicated systems are used in actual power grids. 

\end{document}